\documentclass[prl,twocolumn,showpacs,floatfix,longbibliography]{revtex4-1}
\usepackage{amssymb, amsbsy, amsmath, latexsym, dsfont, array, layout,
graphicx,mathrsfs,color,ulem,bm}

\usepackage{subfigure}
\usepackage{epsfig}
\usepackage{dcolumn}   
\usepackage{bm}        
\usepackage{amssymb,amsmath,amsopn,amsfonts}
\usepackage{colordvi}
\usepackage{color}
\usepackage{multirow}
\usepackage{subeqnarray}
\usepackage{cases}
\usepackage[margin=1in]{geometry}
\usepackage{braket}

\usepackage{dsfont}
\usepackage{wrapfig}

\usepackage{amsmath}
\usepackage{amsfonts}
\usepackage{graphicx}
\usepackage{bbold}
\usepackage{xcolor}

\begin{document}
\title{Conserved quantities in parity-time symmetric systems}
\author{Zhihao Bian$^{1,2}$, Lei Xiao$^{1,3}$, Kunkun Wang$^{1,3}$, Xiang Zhan$^{1,4}$, Franck Assogba Onanga$^5$, Frantisek Ruzicka$^{5,6}$, Wei Yi$^{7,8}$, Yogesh N. Joglekar$^5$, Peng Xue$^{1}$}
\address{$^1$ Beijing Computational Science Research Center, Beijing 100084, China}
\address{$^2$ School of Science, Jiangnan University, Wuxi 214122, China}
\address{$^3$ Department of Physics, Southeast University, Nanjing 211189, China}
\address{$^4$ School of Science, Nanjing University of Science and Technology, Nanjing 210094, China}
\address{$^5$ Department of Physics, Indiana University Purdue University Indianapolis (IUPUI), Indianapolis, Indiana 46202, USA}
\address{$^6$ Institute of Nuclear Physics, Czech Academy of Sciences, Rez 250 68, Czech Republic}
\address{$^7$ Key Laboratory of Quantum Information, University of Science and Technology of China, CAS, Hefei 230026, China}
\address{$^8$ CAS Center For Excellence in Quantum Information and Quantum Physics}

\begin{abstract}
Conserved quantities such as energy or the electric charge of a closed system, or the Runge-Lenz vector in Kepler dynamics are determined by its global, local, or accidental symmetries. They were instrumental to advances such as the prediction of neutrinos in the (inverse) beta decay process and the development of self-consistent approximate methods for isolated or thermal many-body systems. In contrast, little is known about conservation laws and their consequences in open systems. Recently, a special class of these systems, called parity-time ($\mathcal{PT}$) symmetric systems, has been intensely explored for their remarkable properties that are absent in their closed counterparts. A complete characterization and observation of conserved quantities in these systems and their consequences is still lacking. Here we present a complete set of conserved observables for a broad class of $\mathcal{PT}$-symmetric Hamiltonians and experimentally demonstrate their properties using single-photon linear optical circuit. By simulating the dynamics of a four-site system across a fourth-order exceptional point, we measure its four conserved quantities and demonstrate their consequences. Our results spell out non-local conservation laws in non-unitary dynamics and provide key elements that will underpin self-consistent analysis of non-Hermitian quantum many-body systems that are forthcoming.
\end{abstract}
\maketitle

{\it Introduction:---}In Lagrangian dynamics, conservation laws are tied to symmetries of a system through Noether's theorem~\cite{LL1,LL2}. They give rise to global constraints that must be satisfied by approximate methods that are used to model the dynamics. Their influence is ubiquitous in the perturbative, variational, and computational methods for interacting (many-body) systems: only approximations that satisfy conservation laws~\cite{conserving1,conserving2} are physically meaningful and computationally stable. In traditional quantum theory, an observable $O$ is called conserved if it commutes with the Hermitian Hamiltonian $H_0$ of the system. Due to the equivalence between conservation and commutation, a unitary symmetry transformation on the quantum state-space is generated by each conserved observable $O$, thereby reducing complexity of the eigenvalue problem as transformation block-diagonalizes the Hamiltonian. A complete set of conservation laws for a system is obtained by identifying all linearly independent observables that commute with the Hamiltonian. In particular, norm of a quantum state is conserved under the unitary time evolution by any Hermitian Hamiltonian because the operator $O=\mathbb{1}$ trivially commutes with any Hamiltonian.

$\mathcal{PT}$-symmetric systems~\cite{CKR+10,alois2012,feng2014,hossein2014,BSF+14,rotter16,WKP+16,AYF17,HAS+17,WOZ+17,LXZ+17,ZXB+17,WXQ+18,XQW+18,WQX+19,WQXZ+19,XWZ+19,XDW+19,Luo19,WLG+19,NAJM19} are open systems with balanced gain and loss. They are described by a Hamiltonian $H_\mathcal{PT}$ that is invariant under the combined operation of parity and time reversal, and undergoes a time evolution that does not conserve the state norm~\cite{RKM+18}. The spectrum of $H_\mathcal{PT}$ changes from real into complex-conjugate pairs when the gain-loss strength is increased. This $\mathcal{PT}$-symmetry breaking transition occurs at an exceptional point (EP) where eigenvalues and the corresponding eigenmodes coalesce~\cite{kato56}. Non-unitary dynamics of this transition have been observed in classical systems~\cite{CKR+10,alois2012,feng2014,hossein2014,BSF+14,rotter16,WKP+16,AYF17,HAS+17,WOZ+17} and non-interacting quantum systems comprising single photons~\cite{LXZ+17,ZXB+17,WXQ+18,XQW+18,WQX+19,WQXZ+19,XWZ+19,XDW+19}, ultracold atoms~\cite{Luo19}, single spins~\cite{WLG+19}, and superconducting qubits~\cite{NAJM19}.

Complete characterization of conservation laws in such systems is an outstanding question. What are the conserved quantities in $\mathcal{PT}$-symmetric open systems? How do they constrain the (approximate methods used to model the) dynamics, particularly in the $\mathcal{PT}$-broken region where amplifying eigenmodes occur? With an approach inspired by early works on pseudo-Hermiticity~\cite{CDH02,A02,A10}, we conclusively address these questions. For a broad class of $\mathcal{PT}$-symmetric Hamiltonians encompassing all experimentally relevant models, we analytically construct a complete set of linearly independent observables whose expectation values do not change with time. We demonstrate our construction with a four-site $\mathcal{PT}$-symmetric Hamiltonian by encoding the four sites in the path and polarization of a single photon and simulating its non-unitary dynamics. We track four constants of motion across the $\mathcal{PT}$-transition, and demonstrate the consequences of these non-local, conserved quantities through the dynamics of adjacent-site phase differences and conserved-observable eigenstates.

{\it Conserved observables for $H_\mathcal{PT}$:---}Consider a $\mathcal{PT}$-symmetric system described by a $d$-dimensional Hamiltonian $H_\mathcal{PT}$ with an energy scale $J$ ($\hbar=1$). An observable $\hat\eta$ is called an intertwining operator~\cite{A02,A10} for $H_\mathcal{PT}$, if it satisfies
\begin{equation}
\label{eq:eta}
\hat\eta H_\mathcal{PT}=H^\dagger_\mathcal{PT}\hat\eta.
\end{equation}
It follows the non-unitary time evolution operator $G(t)=\exp(-i H_\mathcal{PT}t)$ keeps the observable $\hat\eta$ unchanged, i.e. $G^\dagger(t)\hat\eta G(t)=\hat\eta$. Expectation value of $\hat\eta$ in an arbitrary quantum state is, therefore, a conserved quantity. The equivalence between conservation and commutation breaks down for a non-Hermitian Hamiltonian. In principle, the complete set of conserved observables can be obtained by numerically solving the set of $d^2$ linear equations (\ref{eq:eta})~\cite{MRA+14,choi18}.

Instead, we analytically obtain the complete set of intertwining operators for all $\mathcal{PT}$-symmetric Hamiltonians that are also transpose symmetric, i.e. $H_\mathcal{PT}=H^T_\mathcal{PT}$. This broad class includes all experimentally investigated $\mathcal{PT}$-symmetric systems in the classical~\cite{CKR+10,alois2012,feng2014,hossein2014,BSF+14,rotter16,WKP+16,AYF17,HAS+17,WOZ+17} and quantum~\cite{LXZ+17,ZXB+17,WXQ+18,XQW+18,WQX+19,WQXZ+19,XWZ+19,XDW+19,Luo19,WLG+19,NAJM19} domains, and most of the tight-binding models~\cite{review13}. The recursive construction is as follows: the transpose symmetry implies $\mathcal{T}H_\mathcal{PT}\mathcal{T}=H^\dagger_\mathcal{PT}$, where the time-reversal operator $\mathcal{T}$ is given by complex conjugation. It follows from the $\mathcal{PT}$ symmetry of the Hamiltonian that the parity operator $\mathcal{P}$ is a conserved observable, i.e. $\hat\eta_1=\mathcal{P}$. We then construct a sequence of linearly independent, dimensionless observables $\hat\eta_i=\hat\eta_{i-1} H_\mathcal{PT}/J$ ($i=2,\cdots,d$). The intertwining nature of $\hat\eta_{i-1}$ implies that $\hat\eta_i$ is also an intertwining operator or, equivalently, a conserved observable. This sequence terminates with $\hat{\eta}_d$ because the characteristic equation for $H_\mathcal{PT}$ is a polynomial of order $d$. Thus, $\hat{\eta}_{d+1}$ is a linear combination of lower-order conserved observables. By expressing $\hat\eta$ and $H_\mathcal{PT}$ in the bi-orthogonal eigenbasis of $H_\mathcal{PT}$~\cite{A10}, there are no additional intertwining operators when the spectrum of $H_\mathcal{PT}$ is non-degenerate. Our procedure yields the set of $d$ linearly independent conserved observables for a $d$-dimensional $\mathcal{PT}$-symmetric system. 

\begin{figure*}
\centering
\includegraphics[width=\textwidth]{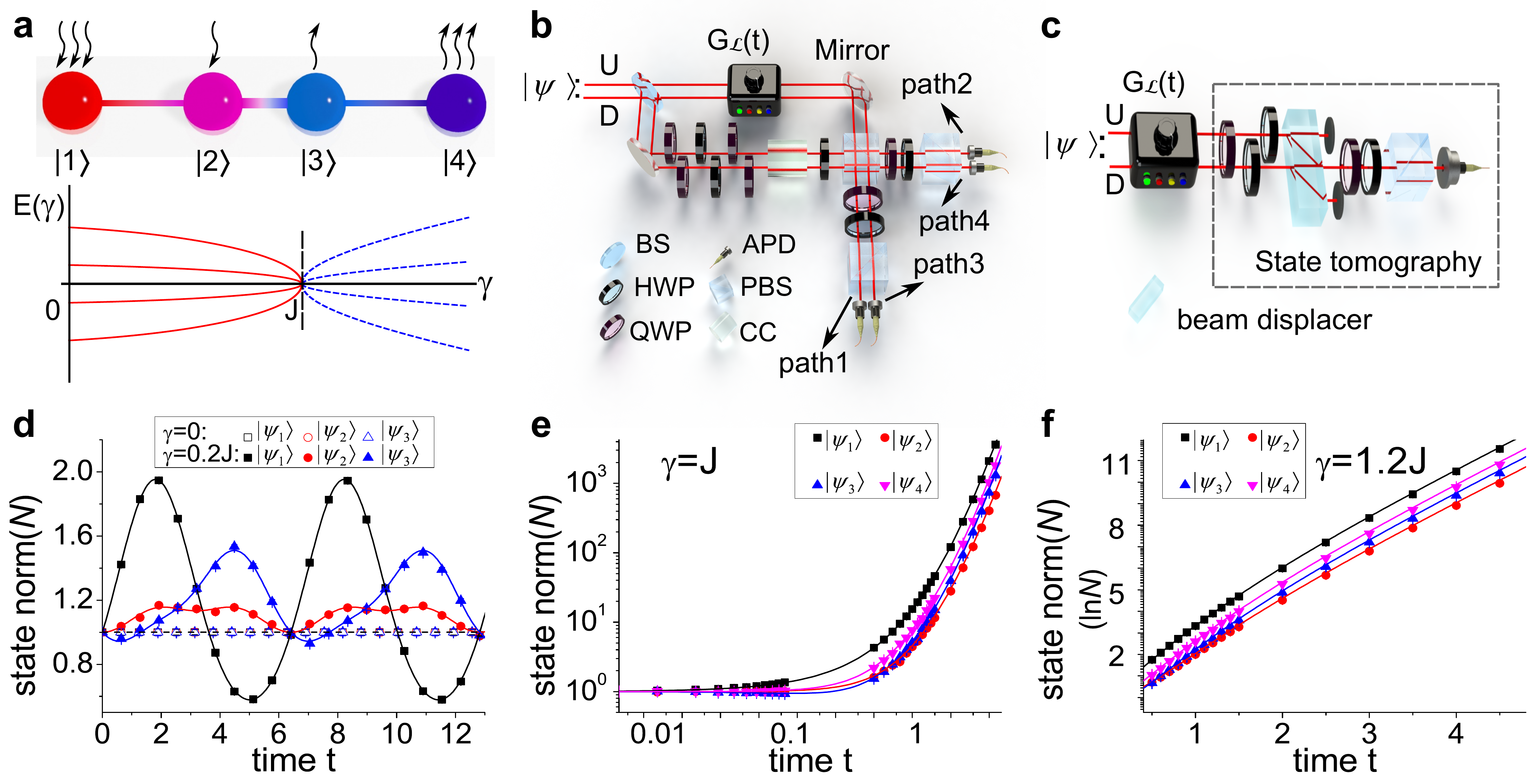}
\caption{Non-unitary dynamics of a four-site $\mathcal{PT}$-symmetric system. {\bf a} Schematic of a four-site system with nearest-neighbor tunneling, gain for sites $|1\rangle$ and $|2\rangle$, and corresponding loss for sites $|4\rangle$ and $|3\rangle$. Spectrum of $H_\mathcal{PT}$. 
{\bf b} Schematic of the optical circuit used to measure the state norm and the inner product between the initial and final states. BS: beam splitter; HWP: half-wave plate; QWP: quarter-wave plate; APD: avalanche photodiode; PBS: polarizing beam splitter; CC: compensated crystal; $G_\mathcal{L}(t)$: lossy time evolution circuit~\cite{sm}. {\bf c} Schematic of quantum-state tomography used to reconstruct the time-evolved state $|\psi(t)\rangle$~\cite{sm}. {\bf d} In the Hermitian limit with $\gamma=0$, measured state norm $N(t)$ is conserved and remains unity (open symbols). In the $\mathcal{PT}$-symmetric region with $\gamma=0.2J$, $N(t)$ oscillates with a period $T(\gamma)$ (filled symbols). {\bf e} At the $\mathcal{PT}$ transition point $\gamma=J$, state norm grows algebraically, $N(t)\propto t^6$. {\bf f} In the $\mathcal{PT}$-symmetry broken region with $\gamma=1.2J$, state norm grows exponentially with time. 
Experimental errors are due to photon-counting statistics; when not shown, error bars are smaller than the symbol size.}
\label{fig:1}
\vspace{-3mm}
\end{figure*}

{\it Non-unitary dynamics of a four-site $\mathcal{PT}$-symmetric system:---}For an experimental demonstration of conserved observables under non-unitary dynamics, we use the Hamiltonian
\begin{align}
\label{eq:hpt}
H_\mathcal{PT}(\gamma)=\frac{1}{2}
\begin{pmatrix}
  3i\gamma & -\sqrt{3}J & 0 & 0 \\
-\sqrt{3}J & i\gamma & -2J & 0 \\
0 & -2J & -i\gamma & -\sqrt{3}J\\
0 & 0 & -\sqrt{3}J & -3i\gamma
 \end{pmatrix},
\end{align}
which is compactly written as $H_\mathcal{PT}(\gamma)=-JS_x+i\gamma S_z$, where $S_x$ and $S_z$ are spin-$3/2$ representations of the SU(2) group~\cite{EUH+08}. It describes a four-site system with mirror-symmetric tunneling, a linear gain-to-loss profile, and the parity operator $\mathcal{P}=\mathrm{antidiag}(1, 1, 1, 1)$. Four equally spaced eigenvalues are given by $\{-3/2,-1/2,+1/2,+3/2\}\sqrt{J^2-\gamma^2}$, which give rise to a fourth-order EP at the $\mathcal{PT}$-breaking threshold $\gamma=J$ (Fig.~\ref{fig:1}a). The single energy gap $\sqrt{J^2-\gamma^2}$ in the spectrum of (\ref{eq:hpt}) leads to dynamics with period $T(\gamma)=2\pi/\sqrt{J^2-\gamma^2}$ for $\gamma< J$.

We encode four sites of the system in the path and polarization degrees of freedom of a single photon, and label them as $|1\rangle=|UH\rangle,|2\rangle=|UV\rangle,|3\rangle=|DH\rangle,|4\rangle=|DV\rangle$. Here $\{|H\rangle,|V\rangle\}$ are horizontal and vertical polarizations, and $\{|U\rangle,|D\rangle\}$ denote upper and lower paths, which undergo gain and loss, respectively. Mapping $H_{\mathcal{PT}}$ into a Hamiltonian with site-selective loss $H_\mathcal{L}(\gamma)=H_\mathcal{PT}(\gamma)-3i\gamma\mathbb{1}/2$, we implement the operator $G_\mathcal{L}(t)=\exp(-iH_{\mathcal{L}}t)$ via a lossy linear optical circuit, which is related to $G(t)$ through $G(t)= G_\mathcal{L}(t)\exp(3\gamma t/2)$~\cite{LXZ+17}. Such a transformation adds an overall gain to experimental measurements, which enables the experimental system with passive $\mathcal{PT}$ symmetry to simulate ideal $\mathcal{PT}$-symmetric models~\cite{LXZ+17,ZXB+17,WXQ+18,XQW+18,WQX+19,WQXZ+19,XWZ+19,XDW+19}. By projecting the time-evolved state $|\psi(t)\rangle=G(t)|\psi(0)\rangle$ onto the site-modes $|k\rangle$ ($k=1,\cdots,4$), time-dependent state norms are obtained (Fig.~\ref{fig:1}b). To probe conserved quantities $\eta_i(t)\equiv\langle\psi(t)|\hat\eta_i|\psi(t)\rangle$, quantum-state tomography is carried out on time-evolved states (Fig.~\ref{fig:1}c). We reconstruct the time evolution over multiple time scales and a wide range of $\gamma$ for four initial states given by $|\psi_1\rangle=|1\rangle,|\psi_2\rangle=(|1\rangle+|2\rangle+|3\rangle+|4\rangle)/2, |\psi_3\rangle=(|1\rangle+\sqrt{2}|4\rangle)/\sqrt{3}, |\psi_4\rangle=(|1\rangle+|4\rangle)/\sqrt{2}$. Conserved quantities $\eta_i(t)$ can also be directly probed via projective measurements~\cite{sm}.

Figures~\ref{fig:1}d-\ref{fig:1}f exemplify the non-unitary dynamics generated by $H_\mathcal{PT}(\gamma)$ through a time-dependent state norm $N(t)=\langle\psi(t)|\psi(t)\rangle$. Constant in the Hermitian limit ($\gamma=0$), $N(t)$ oscillates with a period $T(\gamma)$ in the $\mathcal{PT}$-symmetric region (Fig.~\ref{fig:1}d). State norms do not oscillate around unity, due to the non-othogonality of eigenvectors of the non-Hermitian Hamiltonian $H_\mathcal{PT}(\gamma)$~\cite{sm}. At the fourth-order EP, the state norm $N(t)$ grows algebraically with time as $t^6$ (Fig.~\ref{fig:1}e). Such a scaling is dictated by the order of the exceptional point: at the EP, $H_\mathcal{PT}^4=0$ and the power-series expansion of $G(t)$ terminates at the third order, giving rise to the $t^6$ dependence for the norm. In the $\mathcal{PT}$-symmetry broken region, the measured norm grows exponentially with time (Fig.~\ref{fig:1}f).

\begin{figure*}
\centering
\includegraphics[width=\textwidth]{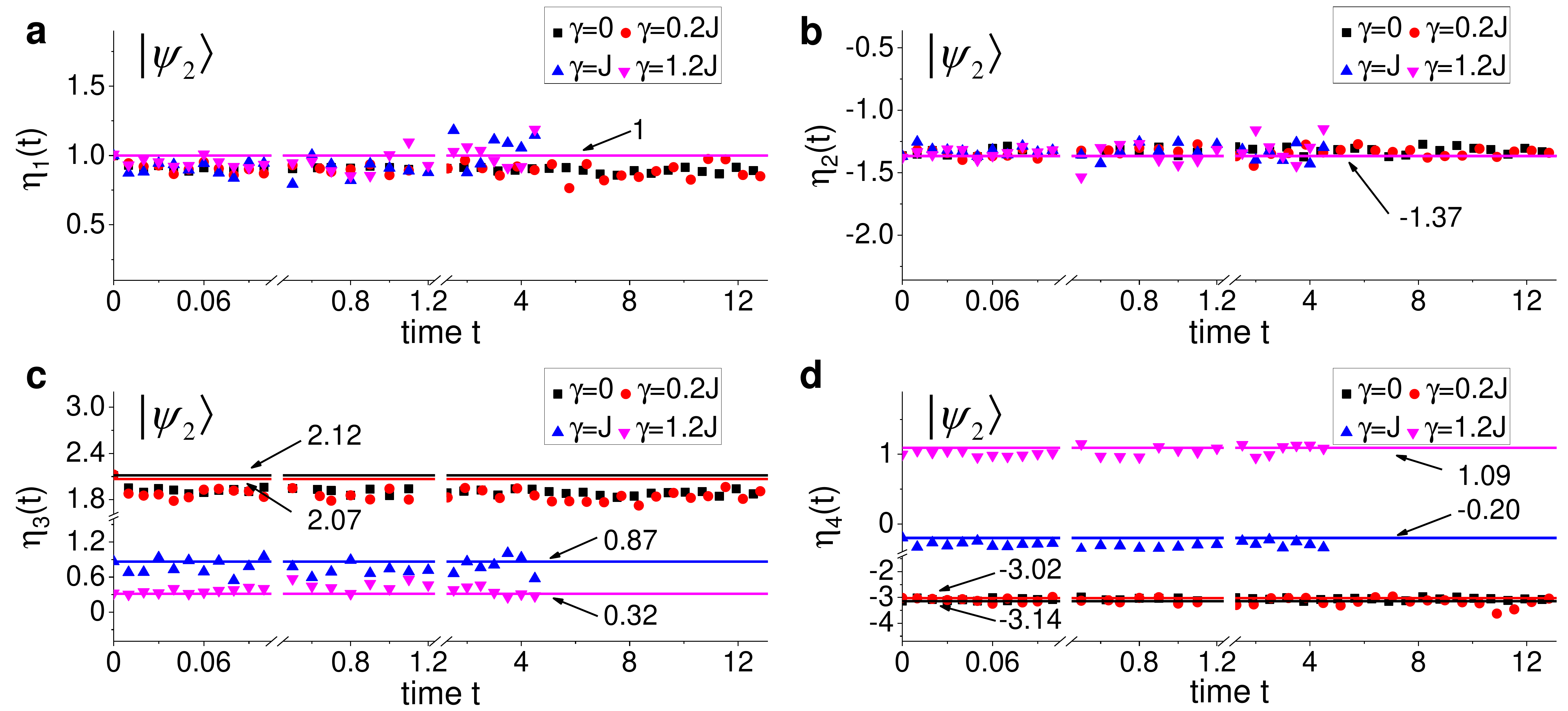}
\caption{Non-local, conserved observables across the $\mathcal{PT}$-transition. {\bf a-d} For a symmetric initial state $|\psi(0)\rangle=|\psi_2\rangle$, measured expectation values $\eta_i(t)$ of the four, dimensionless observables depend on $\gamma$, but remain time invariant. The symmetry of the initial state is responsible for the $\gamma$-independent behavior of the positive $\eta_1(t)$ and negative $\eta_2(t)$.
Error bars 
are smaller than the symbol size and not shown.
Conserved quantities $\eta_i(t)$ for $t<1.2$ and $t>4$ in the unitary and $\mathcal{PT}$-symmetric cases, and that for $t<1.2$ in the $\mathcal{PT}$-broken case (and at the exceptional point) are obtained by direct measurement~\cite{sm}. By contrast, $\eta_i(t)$ for $1.2<t<4$ are measured through state tomography. The experimental results of conserved quantities obtained from both methods are consistent with each other, and match well with theoretical predictions.
}
\label{fig:2}
\vspace{-3mm}
\end{figure*}

{\it Measurement of four non-local conserved quantities:---}For our four-site system, the recursive procedure gives four conserved observables. Expressing the time-evolved state in the site-basis $|\psi(t)\rangle=\sum_{k=1}^4 a_k(t) |k\rangle$, the expectation value of $\hat\eta_1$ is given by~\cite{CDH02,A02,A10}
\begin{equation}
\label{eq:nonlocal}
\eta_1(t)=\sum_{k=1}^4 a^*_{5-k}(t) a_k(t),
\end{equation}
with similar, non-local expressions for the remaining three conserved observables $\hat\eta_2,\hat\eta_3$ and $\hat\eta_4$~\cite{sm}.
Therefore, the global conservation of $\eta_i(t)$ does not translate into local densities that obey continuity equations. This is in stark contrast with the Hermitian or thermal cases where a globally conserved quantity, such as the momentum, gives rise to a local continuity equation involving the momentum-density and the corresponding stress tensor~\cite{LL1}.

Figure~\ref{fig:2} shows measured expectation values $\eta_i(t)$ for the symmetric initial state $|\psi_2\rangle$. While generically dependent on $\gamma$, the expectation values $\eta_i(t)$ remain constant regardless of whether the system is Hermitian ($\gamma=0$), in the $\mathcal{PT}$-symmetric region ($\gamma=0.2J$), at the fourth-order EP ($\gamma=J$), or deep in the $\mathcal{PT}$-symmetry broken region ($\gamma=1.2J$). While $\eta_1(t)$ is positive for all $\gamma$ due to the symmetric nature of $|\psi_2\rangle$, it is not positive-definite over the entire quantum state space. 

Results in Fig.~\ref{fig:2} clearly demonstrate that despite their non-unitary evolution, open systems governed by $\mathcal{PT}$-symmetric Hamiltonians support non-local, conserved quantities. Their constancy at the EP and in the $\mathcal{PT}$-symmetry-broken region, where the state norm increases with time algebraically or exponentially, leads to surprising consequences that we now discuss.

\begin{figure*}
\centering
\includegraphics[width=\textwidth]{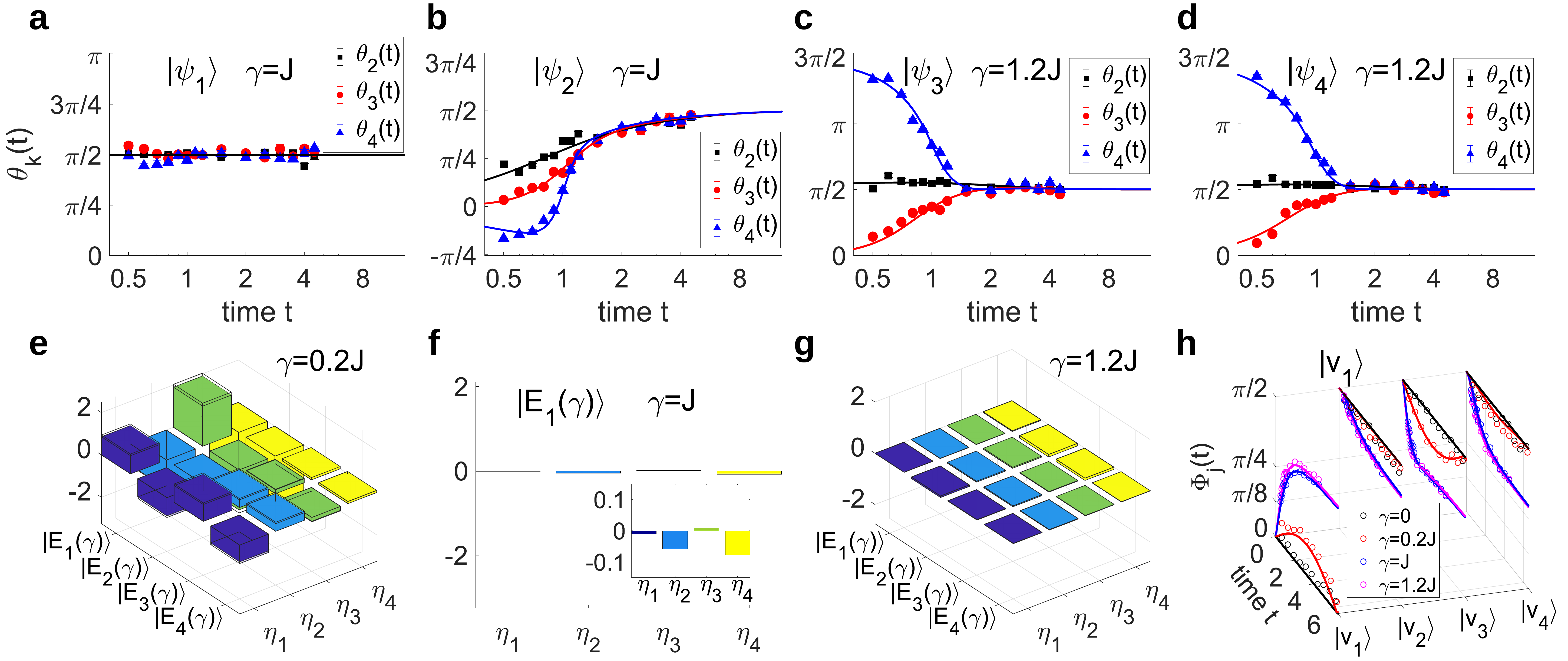}
\caption{Consequences of non-local, conserved observables. {\bf a-\bf b} At the EP, the adjacent-site phase differences $\theta_k(t)$ reach a steady-steady value $\pi/2$ irrespective of the initial state. {\bf c-\bf d} The same phase locking occurs in the $\mathcal{PT}$-symmetry broken region. The diverging state norm and the constancy of $\eta_i(t)$ are responsible for this phenomenon. {\bf e} Conserved quantities $\eta_i$ measured in the $\mathcal{PT}$-symmetric eigenstates $|E_j\rangle$ for $\gamma=0.2J$. {\bf f} At the fourth-order EP, $\gamma=J$, all conserved quantities $\eta_i$ are zero for the sole eigenstate of $H_\mathcal{PT}$. {\bf g} In the $\mathcal{PT}$-symmetry broken region, $\gamma=1.2J$, all $\eta_i$ vanish for each eigenstate $|E_j(\gamma)\rangle$. {\bf h} Measured time-evolution of the angle $\Phi_j(t)$ between $|v_j\rangle$ and $|\psi(t)\rangle$. 
}
\label{fig:3}
\vspace{-3mm}
\end{figure*}

{\it Consequences of conservation laws:---}First, constancy of conserved quantities at the EP or in the $\mathcal{PT}$-symmetry broken region gives rise to a phase-locking phenomenon. Writing the time-evolved state in terms of site amplitudes $r_k(t)$ and phases $\phi_k(t)$ as $|\psi(t)\rangle=\sum_{k=1}^4 r_k e^{i\phi_k}|k\rangle$, we track the phase-difference between adjacent sites, $\theta_k(t)\equiv\phi_k(t)-\phi_{k-1}(t)$ ($k=2,3,4$) for the four initial states. Figures~\ref{fig:3}a-\ref{fig:3}d show that irrespective of the initial state, the phase differences $\theta_k(t)$ reach the steady-state value of $\pi/2$ at the EP (a,b) and in the $\mathcal{PT}$-symmetry-broken region (c,d). Steady-state values are determined by the site distribution of gain and loss, but are independent of the initial state~\cite{YFA18}. The phase locking is the result of an exponential separation between the time dependence of $\langle\psi(t)|\hat{\eta}|\psi(t)\rangle$ and state norm $\langle\psi(t)|\psi(t)\rangle$. Thus, it is also a generic feature of quantum (and classical) systems with site-selective loss, where the $\mathcal{PT}$-symmetry-breaking transition is signaled by the emergence of slowly decaying eigenmodes.

Second, since equivalence between conservation and commutation breaks down for a $\mathcal{PT}$-symmetric system, conserved observables $\hat\eta_i$ and the Hamiltonian $H_\mathcal{PT}$ cannot be simultaneously diagonalized (except at $\gamma=0$). Thus, non-orthogonal eigenmodes $|E_j(\gamma)\rangle$ ($j=1,\cdots,4$) of the Hamiltonian, and orthogonal eigenstates $|v_j\rangle$ of a conserved observable have qualitatively different dynamics.

Figures~\ref{fig:3}e-\ref{fig:3}g show the measured expectation values of the four conserved observables $\hat{\eta}_i$ for different $\gamma$, with the system initialized in different eigenmodes $|E_j(\gamma)\rangle$. The experimental result (colored boxes) is obtained by averaging the measured values $\eta_i(t)$ over all time-points~\cite{sm}; theoretical predications are represented by open boxes. At the fourth-order EP, all expectation values $\eta_i$ for the only eigenstate of $H_\mathcal{PT}$ vanish (Fig.~\ref{fig:3}f). Deep in the $\mathcal{PT}$-symmetry-broken region ($\gamma=1.2J$), $\eta_i$ are zero for each of the four, non-orthogonal eigenmodes $|E_j(\gamma)\rangle$. In a Hermitian system, energy eigenmodes can have multiple, non-zero constants of motion. In contrast, non-local conserved quantities for a $\mathcal{PT}$-symmetric system vanish identically for all eigenmodes that participate in the $\mathcal{PT}$-symmetry breaking transition.

Lastly, in a closed system, a coherent superposition of eigenstates of a conserved quantity cannot be generated from a single, initial eigenstate. In a $\mathcal{PT}$-symmetric system, however, the eigenstates of a conserved observable exhibit nontrivial dynamics. As an example, we consider the dynamics of orthonormal eigenstates $|v_j\rangle$ of a conserved observable $\hat{\eta}_1$. Starting with the initial state $|\psi(0)\rangle=|v_1\rangle$, the experimentally measured time-evolution of the angle $\Phi_j(t)$ between $|v_j\rangle$ and $|\psi(t)\rangle$ is shown in Fig.~\ref{fig:3}h. Fixed, for all times, at zero or $\pi/2$ in the Hermitian limit ($\gamma=0$), angles $\Phi_j(t)$ vary periodically with period $T(\gamma)$ in the $\mathcal{PT}$-symmetric region ($\gamma=0.2J$), but reach steady-state value at the EP ($\gamma=J$) or in the $\mathcal{PT}$-symmetry-broken region ($\gamma=1.2J$). These results also hold for quantum (and classical) systems with mode-selective loss.

{\it Outlook:---}Recent advances have led to the realization of $\mathcal{PT}$ symmetry in minimal quantum systems such as a single spin through Hamiltonian dilation~\cite{WLG+19} and a single superconducting transmon through post-selection~\cite{NAJM19}. These systems are integral to near-term~\cite{CRJ+13} or well-established~\cite{Google} large-scale quantum simulators that will address fundamental, hard questions in strongly correlated, many-body systems. Modeling the dynamics of such systems with $\mathcal{PT}$ symmetry is a challenging open question. Approximate methods such as tensor networks~\cite{orus2014}, the density matrix renormalization group~\cite{dmrg2005}, or the quantum Monte Carlo~\cite{qmc2001}, are based on conservation laws for unitary (or thermal) time evolution, and do not apply to strongly correlated, many-body $\mathcal{PT}$-symmetric systems.

By theoretically revealing and experimentally confirming non-local conservation laws in non-unitary dynamics, our work provides key elements for self-consistent analysis of quantum, many-body, open systems with $\mathcal{PT}$ symmetry, which are forthcoming in the near future. Understanding conserved observables in these systems offers useful insight into quantum dynamics therein, which stimulates the development of new, self-consistent approximation methods for open systems in general.

\begin{acknowledgments} This work has been supported by the Natural Science Foundation of China (Grant Nos. 11674056 and U1930402), the Natural Science Foundation of Jiangsu Province (Grant No. BK20190577), the National Key R\&D Program (Grant Nos. 2016YFA0301700,2017YFA0304100), the National Science Foundation grant DMR-1054020 and the startup funding of Beijing Computational Science Research Center. YJ thanks Andrew Harter and Joshua Feinberg for discussions.
\end{acknowledgments}
%
%
%
%



\end{document}